\documentclass[aps,twocolumn,showpacs]{revtex4}
\usepackage{graphicx}
\usepackage{amssymb}
\usepackage{amsmath}

\begin{document}

\title{$1/f$ noise and avalanche scaling in plastic deformation}

\author{L. Laurson and  M.\ J.\ Alava}

\affiliation{Helsinki University of Technology, Laboratory of Physics,\\
P.O.Box 1100, FIN-02015 HUT, Finland }

\begin{abstract}
  \noindent
We study the intermittency and noise of dislocation systems undergoing
shear deformation. Simulations
of a simple two-dimensional discrete dislocation dynamics model indicate
that the deformation rate
exhibits a power spectrum scaling of the type $1/f^{\alpha}$.
The noise exponent is far away from a Lorentzian, with
$\alpha \approx 1.5$. This result is directly related
to the way the durations of avalanches of plastic deformation activity
scale with their size. 
\end{abstract}
\maketitle

\date{\today}

\section{Introduction}

Traditionally, the plastic deformation of solids has been thought of as a 
smooth process, so that the fluctuations can be neglected
on large enough scales as the behavior becomes homogeneous.
Contrary to this old paradigm, very recent evidence from experiments 
points out clearly that plastic deformation proceeds via bursts of dislocation 
activity \cite{miguel,dimiduk,weiss,marsan,zaiser,weiss_jcp,zaiserreview}. 
This activity, due to the long-range interactions, should have
extended spatial and temporal correlations. Supporting this
new picture, a number of simulation studies of simplified models 
\cite{miguel,miguel2,koslowski,morettizaiser_jstat} have demonstrated 
that such bursts or avalanches can be characterized by, apparently,
scale-free size distributions. 

These advances bring plasticity close to a multitude of systems exhibiting
``crackling noise'' \cite {sethna}. This is a generic property of many 
non-equilibrium systems, such as earthquakes \cite{gr}, Barkhausen noise in 
ferromagnets  \cite{durin}, fracture \cite{ciliberto,salminen},
vortex avalanches in superconductors \cite{field} and 
martensitic shape memory alloys \cite{vives} to name a few.
The common feature among these are metastability and
avalanches of activity that intercept quiescent periods.
The realization that plastic deformation has rich internal
dynamics is of interest both for physics and materials science.
Dislocations provide an interesting example of
the physics of systems with long-range interactions
competing with pinning from obstacles such as impurities,
grain boundaries, and other dislocations  \cite{zaiserreview,morettizaiser}  
(``forest hardening'' being a well-known example of the ensuing phenomena).

Such avalanching systems are most simply studied by
the one-dimensional activity time series $V(t)$. In plasticity, $V$
could be the global strain rate, or the acoustic emission activity 
during deformation, or in a constant strain rate experiment the
shear stress.
In this work we consider the temporal characterization of
dislocation activity and show that its power spectrum 
(PS) exhibits what is known as $1/f$-noise, 
$P(f)\sim 1/f^{\alpha}$. This ``flicker'' noise is a
well-known but relatively little understood phenomenon
in various fields of science \cite{1/f}.
Then, we relate the scaling exponent $\alpha$ to the
characteristic scaling of the avalanches. This has to be done with care
due to the low-level background activity included in $V$.
At the end, we are able to demonstrate that the noise of irreversible 
deformation may follow generic, experimentally verifiable scaling laws.
We use a simple two-dimensional (2d) discrete dislocation 
dynamics (DDD) model \cite{miguel2}, and study its behavior
in the steady-state regime with a constant shear stress
and a small (on the average constant) deformation rate.
While the statistics of avalanche sizes in this kind of models has
been established to be of power law type \cite{miguel,miguel2}, 
no temporal scaling analysis has been presented so far.

The PS $P(f)$ of a time series $V(t)$ is the
absolute square of the Fourier transform of $V(t)$, 
\begin{equation}
  P(f) \sim \left|\int e^{i2\pi ft}V(t)dt\right|^2.
\end{equation}
Since  the time-time correlation function and the PS are related,
the latter is a measure of temporal correlations in the system.
$\alpha$-values less than two indicate the presence of 
complex time correlations. Recently it has been
realized that under certain fairly general conditions the exponent $\alpha$ 
can be derived from the scaling exponents 
characterizing the avalanche distributions, the
examples ranging from models of Barkhausen noise \cite{KS}, to
self-organized criticality \cite{oma} to fluid
invasion into disordered media \cite{rost}.
The intermittency is easier to understand
in toy models as sandpile models of self-organized criticality 
\cite{btw}, where $V(t)$ drops to zero between avalanches.
In most experiments and more realistic model systems - such as here -
there is a background (from noise, or processes that co-exist with
the avalanches). Also, finite drive rates can lead to merging of avalanches,
thus to problems in characterizing the underlying activity.

The structure of the paper is as follows:
In Sec. II we introduce the 2d DDD model, and in Sec. III the 
scaling relation relating the power spectrum exponent to the 
avalanche statistics is presented. Results of the numerical
simulations are presented in Sec. IV. Section V finishes the paper 
with conclusions.

\section{DDD model}

Despite its simplicity, the DDD model we study here has been shown to
capture many features of realistic plasticity, such as avalanches
with scale-free size distributions \cite{miguel,miguel2} as well as 
the Andrade creep law \cite{miguelprl}. 
It is a two-dimensional model representing a cross section of a crystal
with single-slip geometry. The dislocations are assumed to be
straight parallel edge dislocations, with the dislocation lines oriented along
the $z$ axis. They glide along directions 
parallel to their Burgers vectors ${\bf b}=\pm b {\bf u}_x$, with ${\bf u}_x$
the unit vector in the $x$-direction. It is thus sufficient to consider
a two-dimensional system (i.e. the $xy$ plane) with $N$ pointlike edge 
dislocations gliding in the $x$-direction. Equal numbers of dislocations
with positive and negative Burgers vectors are assumed. For simplicity 
other processes contributing to dislocation motion such as dislocation 
climb are not considered.  

An essential feature of the model is that the dislocations interact
with each other through their anisotropic long-range stress fields
\begin{equation}
\label{stress}
\sigma_s({\bf r}) = Db\frac{x(x^2-y^2)}{(x^2+y^2)^2},
\end{equation} 
where $D=\mu/2\pi(1-\nu)$, with $\mu$ the shear modulus and $\nu$
the Poisson ratio of the material. We assume overdamped dynamics 
with a linear force-velocity relationship, giving rise to the
equations of motion of the form 
\begin{equation}
\label{eom}
\frac{\chi_d^{-1}v_n}{b} = s_n b \left[ \sum_{m \neq n} s_m 
\sigma_s({\bf r}_{nm}) + \sigma \right],
\end{equation}
where $v_n$ is the velocity of the $n$'th dislocation,  
$\chi_d$ is
the dislocation mobility, $s_n$ refers to the sign of the Burgers
vector of the $n$'th dislocation and $\sigma$ is the external shear
stress experienced by the dislocation. These equations are put into
a dimensionless form by measuring lengths in units of $b$, times
in units of $1/(\chi_d D b)$ and stresses in units of $D$. Due to the
long range nature of the dislocation-dislocation interaction, periodic
boundary conditions are implemented with an infinite sum of images 
\cite{hirth}. The positions ${\bf r}_n$, $n=1,\dots,N$, of the $N$ 
dislocations as a function of time are computed by integrating the 
equations of motion (\ref{eom}) numerically by using an adaptive step 
size fifth-order Runge-Kutta algorithm.

For small distances, Eq. (\ref{stress}) ceases to be valid. 
Thus for distances smaller than $2b$ from the dislocation, we set 
the dislocation stress field to zero. This procedure removes the 
unphysical singularity in Eq. (\ref{stress}). Furthermore, when the 
distance $r_{nm}$ between two dislocations with Burgers vectors of 
opposite sign gets small, i.e. $r_{nm}<2b$, we employ a 
phenomenological annihilation reaction by simply removing them from 
the system. To compensate, and to include dislocation multiplication 
(as in real plasticity through Frank-Read sources), we introduce a 
mechanism to create new dislocations. The system is split into smaller 
cells and then one monitors both the local stress and the number of 
pinned  dislocations (i.e. those moving slower than a threshold velocity)
in each cell. Given pinned dislocations and that the local stress 
exceeds a threshold value, a dislocation pair with 
opposite Burgers vectors is generated with a probability
proportional to the absolute local stress value. This means that
on the average it takes a finite time for the source to create a
new dislocation pair, as is the case with real dislocation sources. 
The two new dislocations of opposite sign are inserted into random 
locations inside the neighboring cells of the one containing the 
activated source. This is done with the constraint that their 
combined stress field must decrease the magnitude of the local stress 
at the source location, while elsewhere in the system further dislocation
activity may be triggered, resulting in a correlated sequence of 
dislocation activity.

\section{Scaling of the power spectrum}

Consider a bursty time series $V(t)$ consisting of temporally  
separated avalanches. The usual definition of an avalanche is a 
connected sequence of values of $V(t)$ exceeding some threshold 
value $V_{th}$, to e.g. subtract uncorrelated background noise. 
If an avalanche starts at $t=0$ and ends at $t=T$, the 
size $s$ of an avalanche of duration $T$ is defined as 
$s(T)=\int_0^T[V(t)-V_{th}]dt$. Assume that the average avalanche 
size $\langle s(T) \rangle$ of avalanches of a given duration $T$ 
scales as a power law of the duration,
\begin{equation}
  \label{sT}
  \langle s(T) \rangle \sim T^{\gamma_{st}},
\end{equation}
and that the avalanches are self-similar so that the averaged
avalanche shapes $V(T,t)$ of avalanches of different durations $T$
can be collapsed onto a single curve by using the ansatz
\begin{equation}
  \label{shape_ansatz}
  V(T,t) = T^{\gamma_{st}-1} f_{shape}(t/T).
\end{equation}
Here, $f_{shape}(x)$ is a scaling function. The total energy is 
obtained as the $\theta=0$ component of the stationary time-time 
correlation function, defined by $C(\theta) = \int V(t)V(t+\theta)dt$.
By cosine transforming the time-time correlation function $C(\theta|s)$ 
of avalanches of a given size $s$, one obtains the scaling form 
\begin{equation}
\label{energy}
E(f|s) = s^2g_E(f^{\gamma_{st}}s)
\end{equation} 
for the corresponding energy
spectrum. The scaling of the total power spectrum then follows
by averaging $E(f|s)$ over the avalanche size probability 
distribution $D_s(s)$, e.g. a power law 
$D_s(s) \sim s^{-\tau_s}$ cut off at $s=s^*$, so that
\begin{equation} 
P(f) = f^{-\gamma_{st}(3-\tau_s)} \int^{s^* f^{\gamma_{st}}} 
dx x^{2-\tau_s} g_E(x).
\end{equation}
The result will depend 
on the value of $\tau_s$ as well as on the form of the scaling function
$g_E(x)$ \cite{KS,oma}. In the case $g_E(x) \sim 1/x$ \cite{KS,oma} 
and $\tau_s < 2$ (for the DDD models at hand, the latter condition
seems to be fulfilled, with $\tau_s \approx 1.6$ \cite{miguel2}) the 
power spectrum scales as
\begin{equation}
  \label{tulos}
  P(f) \sim f^{-\gamma_{st}}.
\end{equation}
This indicates a scaling relation $\alpha = \gamma_{st}$, which 
possibility we next check with simulations.

\section{Results}

A natural
choice for the time series is the ``collective
velocity'' of the dislocations, $V(t)=\sum_i |v_i|$. It is proportional
to the energy dissipated per unit time by the dislocation system. 
Thus the statistical properties of $V(t)$ could be related to
the acoustic emission (AE) statistics, as suggested by a number of authors,
see Ref. \cite{zaiserreview}. Another 
possibility is $V_s(t)=\sum_i b_i v_i$, i.e. the global strain
rate. We mainly present results here for the first choice.

\begin{center}
  \begin{figure}[ht]
    \center
    \includegraphics[width=9cm
    ]{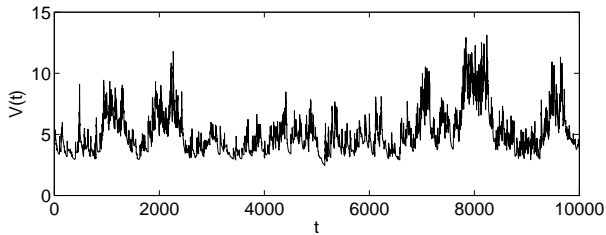}
    \caption{An example of the time series $V(t)=\sum_i |v_i|$ from
      a system of linear size $L=300b$. $\sigma = 0.03$. Time is reported
      in units of $1/(\chi_d Db)$ and $V$ in units of $\chi_d Db^2$. In the 
      steady state, the system contains on the average about 600 
      dislocations.}
    \label{signal}
  \end{figure}
\end{center}

The numerical simulations are started from a random initial configuration
of $N_0$ dislocations with equal numbers of positive and negative Burgers
vectors (here $L=200b$ and $L=300b$,
with $N_0=600$ and $N_0=1000$, respectively). The system is first let to
relax in the absence of external stress, until it reaches a metastable
arrangement with $N<N_0$.
Then a small constant external stress is applied and the evolution of the
system is monitored.
In the absence of dislocation multiplication, the system undergoes a 
``jamming transition'' at a critical value $\sigma_c$ of the external 
stress \cite{miguelprl}.
For $\sigma=\sigma_c$, the strain rate decays as a power law in time, with the
Andrade's power law creep exponent close to $-2/3$.
For $\sigma>\sigma_c \approx 0.01$, 
the system crosses over to a constant strain rate
regime (linear creep) which we study next.

With the dislocation multiplication turned on, the system displays
intermittent avalanche-like bursts of dislocation activity.
The number of dislocations fluctuates around $N \approx 300$ for
$L=200b$ and $N \approx 600$ for $L=300b$. Fig. \ref{signal} shows that
$V(t)$ displays intermittent behavior, consisting of avalanches. These are
defined as a connected sequence of $V(t)$-values exceeding a threshold 
value $V_{th}$ and have a wide range of sizes. Here, we monitor $V(t)$ 
for external stress values close to but above the critical value $\sigma_c$.  

\begin{center}
  \begin{figure}[ht]
    \center
    \includegraphics[width=8cm
    ]{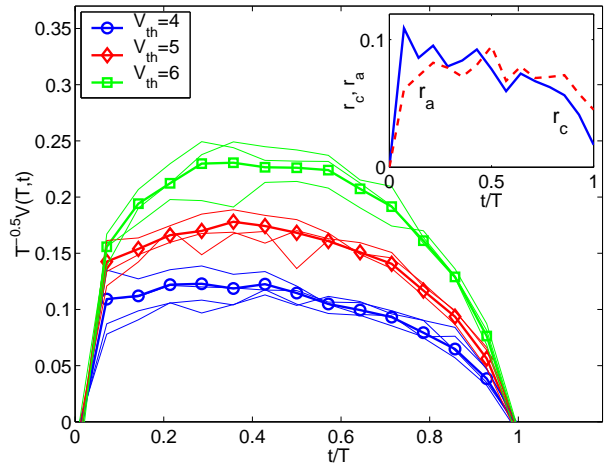}
    \caption{(Color online) The average scaled avalanche shape for different 
      threshold values $V_{th}$ for $\sigma=0.03$ and $L=300b$. For small 
      threshold values the avalanches are asymmetric with the 
      asymmetry decreasing with $V_{th}$.
      The thick lines with circular symbols correspond
      to an average over of the order of $10^{3}$ avalanches, which
      have been scaled according to Eq.
      (\ref{shape_ansatz}) before averaging. The thin solid lines correspond
      to avalanches of different limited duration ranges. The inset
      displays the average creation and annihilation rates 
      ($r_c$ and $r_a$, respectively) of dislocations during an 
      avalanche, with $V_{th}=5$.
    }
    \label{shape}
  \end{figure}
\end{center}

\begin{center}
  \begin{figure}[ht]
    \center
    \includegraphics[width=8cm
    ]{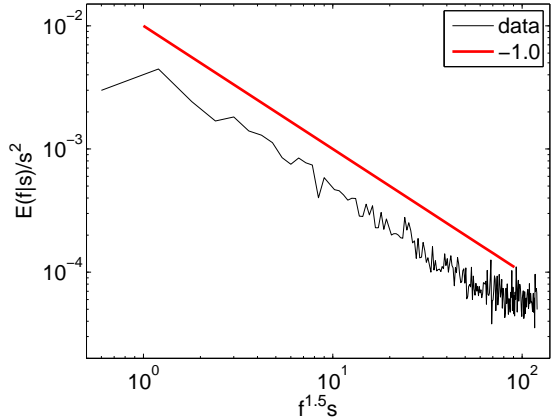}
    \caption{(Color online) The scaled energy spectrum of dislocation 
      avalanches, showing the form of the scaling function $g_E(x)$, 
      which is consistent with the $g_E(x) \sim 1/x$ behavior, indicated
      by the thick solid line. The data has been obtained
      by computing the energy spectrum $E(f|s)$ of each avalanche for 
      $\sigma=0.03$ and $L=300b$, normalizing by $s^2$, and averaging over 
      all avalanches longer than a threshold duration.}
    \label{1_per_x}
  \end{figure}
\end{center}

Fig. \ref{shape} exhibits the average avalanche shape (one needs
a minimum duration, of the order of 10, for scaling), for varying
$V_{th}$.
They appear to agree with Eq. (\ref{shape_ansatz}).
The data has been obtained 
by scaling all individual avalanches in a given duration range according to
Eq. (\ref{shape_ansatz}).
The threshold value $V_{th}$ does not have any effect 
on the $s(T)$ statistics (given that $V_{th}$ is high enough 
such that a ``noise level'' corresponding to {\em incoherent motion} of 
dislocations is not taken into account), but
the avalanche shapes appear to depend
on $V_{th}$. For small $V_{th}$-values, avalanches are clearly asymmetric
to the left (in agreement with experiments \cite{weiss3}), but
become more symmetric upon increasing $V_{th}$.
This asymmetry is also manifest in the creation rate of dislocations during
an avalanche (inset of Fig. \ref{shape}).
In Fig. \ref{1_per_x} we show the energy spectra computed using the Lomb
periodogram, scaled according to Eq. (\ref{energy}).
The behavior of the scaling function $g_E(x)$ indicates that
$g_E(x) \sim 1/x$.

\begin{center}
  \begin{figure}[ht]
    \center
    \includegraphics[width=9cm
    ]{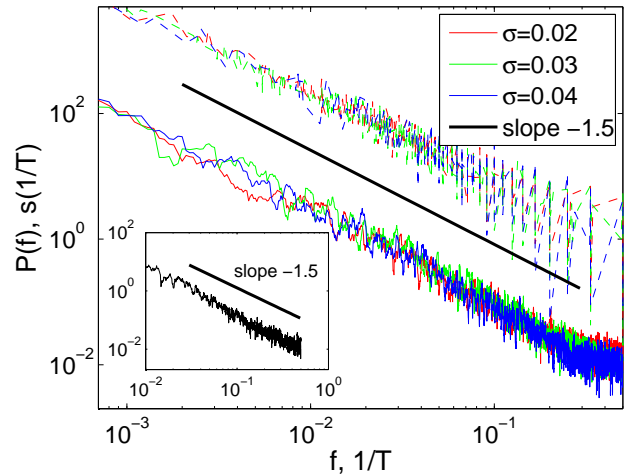}
    \caption{(Color online) A comparison of the scaling of the power 
      spectra (solid lines) with that of $\langle s(1/T) \rangle$ (dashed 
      lines), for a system of linear size $L=300b$ and various 
      $\sigma$-values. The thick solid line corresponds to 
      $\alpha=\gamma_{st}=1.5$. The inset shows the power spectrum of 
      the strain rate, for $\sigma=0.04$. The results are averaged over
      several random initial configurations. Both $f$ and $1/T$ are reported 
      in units of $\chi_d Db$.}
    \label{svst}
  \end{figure}
\end{center}

\begin{center}
  \begin{figure}[ht]
    \center
    \includegraphics[width=9cm
    ]{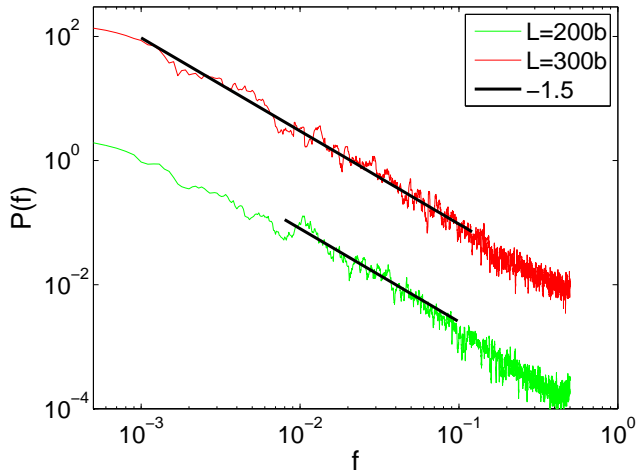}
    \caption{(Color online) Power spectra for two different 
      system sizes, $L=200b$ and $L=300b$, corresponding to on 
      the average 300 and 600 dislocations in the steady state. The 
      extension of the scaling region of the PS is seen to 
      increase with $L$. The results are averaged over
      several random initial configurations. The frequency $f$ is reported 
      in units of $\chi_d Db$.}
    \label{ps_L}
  \end{figure}
\end{center}

The main result is shown in  Fig. \ref{svst}, where we compare the scaling of 
the total power spectrum, from $V(t)$, with that of
$\langle s(T) \rangle$. We observe a PS of the form
$P(f) \sim f^{-\gamma_{st}}$ spanning almost two decades, 
with $\gamma_{st} \approx 1.5$. Thus the PS of dislocation activity
is related here to the intrinsic scaling of the avalanches.
The extension of the scaling regime increases
with $L$, the system size (Fig. \ref{ps_L}), with a cut-off frequency
$f_{cut-off}$ roughly equal to the system size dependent inverse duration 
of the longest avalanche, $f_{cut-off}\sim T_{max}^{-1}(L)$. The absence 
of scaling for the very highest frequencies ($f>0.1$)
is due to a finite crossover time, before the avalanches have a 
self-similar structure. Similarly to the exponent of the avalanche size
distribution \cite{miguel}, the exponent of the power spectrum is 
interestingly insensitive to the value of the external stress. 
In the inset of Fig. \ref{svst} we consider the PS of the 
total strain rate. While the scaling region appears somewhat narrower in 
this case, the exponent $\alpha$ is observed to be unchanged from $1.5$.

These results are largely independent of the details of the model, such
as the threshold value for the local stress to create new dislocations.
We have checked that the same results can be recovered even in the case
with no dislocation multiplication, where the avalanches are due to
different threshold mechanisms, such as unpinning of dislocation dipoles.
As this approach suffers from the fact that simulations are computationally
more demanding due to the longer time scales one must reach, we have here
restricted ourselves to considering only the case where dislocation 
multiplication does occur.

Knowing that the the avalanches are self-affine one can also compute
$\gamma_{st}$ from the other known exponents.
Consider the probability $D_V(V|s)$ that a value $V$ will occur at some
point during an avalanche of size $s$. Assuming scaling and 
requiring normalization, $\int_0^{\infty} D_V(V|s) dV = 1$, as well as 
$\langle V \rangle = s/T \sim s^{1-1/\gamma_{st}}$, one obtains 
(see also \cite{KS}) 
\begin{equation}
D_V(V|s) = V^{-1}f_V(Vs^{1/\gamma_{st}-1}),
\end{equation}
where $f_V(x)$ is a scaling function. The total probability $D_V(V)$ is then
obtained by integrating $D_V(V|s)$ over the avalanche size 
distribution $D_s(s) \sim s^{-\tau_s}$, giving rise to
$D_V(V) \sim  V^{-(\frac{\gamma_{st}\tau_s-1}{\gamma_{st}-1})}
\equiv V^{-\tau_V}$.
Miguel et al. \cite{miguel} studied the statistics of the quantity 
$E = (\sum_i |v_i|)^2$, and found that its distribution 
behaves like $D_E(E) \sim E^{-\tau_E}$,
with $\tau_E \approx 1.8$. From $\tau_E = \frac{1}{2}(\tau_V + 1)$ 
one may solve for $\gamma_{st}$, resulting in (with $\tau_s \approx 1.6$
\cite{miguel2}) 
\begin{equation}
\label{gammaon}
\gamma_{st} = \frac{2\tau_{E}-2}{2\tau_{E}-\tau_s-1} \approx 1.6,
\end{equation}
in reasonable agreement with our results above.

\section{Conclusions}

In this work, we have demonstrated that a simple two-dimensional
dislocation system exhibits $1/f^\alpha$-noise, so that the relation 
$\alpha=\gamma_{st}$ can be explained by the scaling
properties of the avalanche dynamics. 
We have also considered the PS of the global strain rate, 
with similar conclusions. These findings lend themselves
to experimental tests, e.g. as
in the creep experiments of ice single crystals \cite {weiss}, 
in experiments on deforming metallic single (micro)crystals 
\cite{weiss2,dimiduk}, and could also be considered in experiments on 
colloidal crystals \cite{pertsinidis}. 
In AE experiments performed with three-dimensional 
systems, values consistent with mean field exponents
have been reported \cite{weiss,weiss2}, suggesting $\alpha=\gamma_{st}=2$
(see Eq. (\ref{gammaon})). 
It is thus possible that to get quantitative agreement with experiments, 
the full three-dimensional problem
should be studied, e.g. by means of simulations of a three-dimensional 
dislocation dynamics model. Other issues not considered in the present study
include the possible relevance of quenched disorder e.g. in the form
of forest dislocations, which would provide strong pinning centers
to resist dislocation motion. One should note, however, that also
the present model contains a pinning force landscape generated by
the dislocations themselves \cite{miguel}.

In polycrystals, avalanches interact with
the grain boundaries, which will probably lead to
a size-dependent avalanche shape \cite{richeton}. This means that Eq. 
(\ref{shape_ansatz}) is no longer directly applicable, presenting
an interesting theoretical question. Our theory assumes
negligible correlations between avalanches. In experiments, dislocation
avalanches have been found to exhibit a tendency to cluster in time, 
such that a ``mainshock'' is typically followed by a sequence of few 
``aftershocks'' \cite{weiss,dimiduk}. While 
this may modify the low frequency part of the PS, the high frequency 
part, corresponding to correlations within individual avalanches, is 
still expected to scale according to Eq. (\ref{tulos}).

Finally we note that in addition to materials with crystalline 
structure, one field where the use of spectral tools and the 
study of avalanches might be used to characterize the spatio-temporal 
behavior is the plasticity of non-crystalline media, e.g. amorphous 
glasses, where the localization and intermittency of plastic events has 
been recently demonstrated in simulations \cite{lemaitre,tanguy}.

{\bf Acknowledgments}
The Center of Excellence program of the Academy of Finland 
is thanked for financial support.

\end{document}